\documentclass[twoside,11pt]{article}
\usepackage{a4wide}
\usepackage[tbtags]{amstex}
\begin{document}
\renewcommand{\thefootnote}{\arabic{footnote}}
\newcommand{\mc}{\mathcal}
\newcommand{\beq}{\begin{equation}}
\newcommand{\eeq}{\end{equation}}
\newcommand{\beqa}{\begin{eqnarray}}
\newcommand{\eeqa}{\end{eqnarray}}
\newcommand{\beqar}{\begin{eqnarray*}}
\newcommand{\eeqar}{\end{eqnarray*}}
\newcommand{\hra}{\hookrightarrow}
\newcommand{\ul}{\underline}
\newcommand{\df}{\stackrel{def}{=}}
\newcommand{\Llra}{\Longleftrightarrow}
\newcommand{\lra}{\longrightarrow}
\numberwithin{equation}{section}
\def\msr{{\rm I\!R}} 
\def\msn{{\rm I\!N}} 
\def\msc{{\rm C\!\!\!\!\!\;}{\sf I}\;} 
\def\msm{{\rm I\!M}}
\def\msh{{\rm I\!H}}
\def\msk{{\rm I\!K}}
\def\msp{{\rm I\!P}}
\def\mse{{\rm I\!E}}

\makeatletter
\title{Path Integral Quantization of the Poisson-Sigma Model}
\author{Allen C. Hirshfeld\footnote{hirsh@hal1.physik.uni-dortmund.de}\, and 
Thomas Schwarzweller\footnote{thomas@doom.physik.uni-dortmund.de}\\
Fachbereich Physik\\
Universit\"at Dortmund\\
44221 Dortmund}

\maketitle

\begin{abstract}
{\small\sl  We apply the antifield quantization method of Batalin and Vilkovisky to the calculation of 
the path integral for the Poisson-Sigma model in a general gauge. For a linear Poisson 
structure the model reduces to a nonabelian gauge theory, and we obtain the formula for the 
partition function of two-dimensional Yang-Mills theory for closed oriented 
two-dimensional manifolds.}
\end{abstract}
\section{Introduction}
The Poisson-Sigma model associates to any Poisson structure on a finite-dimensional manifold a 
two-dimensional field theory \cite{SS1}. Choosing different Poisson structures leads to 
specific models which include most of the topological and semi-topological field theories 
which have been of interest in recent years. These include gravity models, non-abelian 
gauge theories and the Wess-Zumino-Witten model. Under some natural restrictions these 
models are completely integrable.

When the two-dimensional spacetime is a cylinder the Poisson-Sigma model can 
be quantized with the canonical quantization procedure \cite{SS2}.  In this paper 
we use the path integral method in order to perform the quantization for 
arbitrary world sheet topologies.

In the language of gauge theories the models considered here involve an open 
gauge algebra, i.e. the algebra closes only on-shell. In such cases the 
Faddeev-Popov method of path integral quantization fails. Quantization 
procedures which rely on the BRST symmetry of the extended action are in 
principle more powerful \cite{H}.  We find that for our purposes the antifield formalism 
of Batalin and Vilkovisky \cite{BV} is
the most effective quantization method.
 
The path integral approach for the Poisson-Sigma model was first discussed 
in a preliminary way by Schaller and Strobl \cite{SS2}.  In a recent paper Cattaneo 
and Felder \cite{CF} use the perturbation expansion of the path integral in the 
covariant gauge to elucidate Kontsevich's formula for the deformation 
quantization of the algebra of functions on a Poisson manifold \cite{KO}. Kummer, Liebl and Vassilevich have investigated the special case of 2d dilaton gravity in 
the temporal gauge, and they have calculated the generating functional using BRST 
methods \cite{KU1}. In further work they have studied the coupling to matter fields \cite{KU2}.
We present here a complete and general derivation of the partition function for the 
Poisson-Sigma model for an arbitrary gauge. In the study of
the Chern-Simons topological gauge theory it is known that different choices
of the gauge-fixing function lead to different integral representations of the
associated knot invariants: the light-cone gauge leads to the universal
Kontsevich integral \cite{LP1} , the Landau gauge to the covariant integrals 
of 
Bott and Taubes \cite{BOT},\cite{HS} and the temporal gauge to the singular integrals 
studied by Labastida and P\'{e}rez \cite{LP2}. The relation between these various representations is at 
present not well-understood, in particular no one has been able to reproduce 
the necessary Kontsevich normalization factor from a path integral approach. 
We hope that the techniques developed here may be helpful in that context. 
 
Our paper is structured as follows. In Section 2 we give a concise but self-contained review 
of the Batalin-Vilkovisky quantization procedure for gauge theories.  In Section 3 we apply 
the method to the Poisson-Sigma model. In Section 4 we show how the general model reduces
under certain circumstances to the more familiar Yang-Mills case, and that we reproduce in 
this case the formula for the partition function \cite{BT}.
Section 5 contains our conclusions and an outlook for further research.

\section{The Antifield Formalism for the Quantization\\
of Gauge Theories}
\label{antifield}

\subsection{The Structure Equations of Gauge Theories}

The Batalin-Vilkovisky formalism has a beautiful geometric interpretation, 
first discovered by Witten \cite{Wi1}, and recently described in the paper of 
Alexandrov et. al. \cite{AKSZ}. Here we are not concerned with these 
aspects; we just want to show how the formalism can be applied. We restrict 
our account to irreducible dynamical systems. For further details we refer to 
the recent review by Gomis, Paris and Stuart \cite{GPS}.

We consider a system whose dynamics is governed by a classical action 
$\mc{S}_{0}[\phi^{i}]$ which depends on the fields 
$\phi^{i}(x),\,\,\,i=1,\ldots,n.$  In the following we shall use a compact 
notation in which the multi-index $i$ may denote the various fields involved,
the discrete indices on which they may depend, and the dependence 
on the spacetime variables as well.  The generalized summation convention 
then means that a repeated index may denote not only a sum over discrete 
variables, but also integration over the spacetime variables. 
$\epsilon_i=\epsilon(\phi^{i})$ will denote the Grassman parity of the 
fields $\phi^{i}$.  Fields with $\epsilon_i=0$ are called bosonic, fields 
with  $\epsilon_i=1$ fermionic. The graded commutation rule is
\begin{equation}
\phi^{i}(x)\phi^{j}(y)=(-1)^{\epsilon_{i}\epsilon_{j}}{\phi}^{j}(y){\phi}^{i}(x).
\end{equation}

For a gauge theory the action is invariant under a 
set of $m$ gauge transformations with infinitesimal form 
\begin{equation}
\delta\phi^i=R^i_\alpha \,\varepsilon^\alpha,
\;\;\;\;\alpha=1\,\,{\rm or}\,\, 2\,\, {\rm or}\ldots m.
\end{equation}
This is compact notation for
\begin{eqnarray}
\delta\phi^{i}(x) 
&=& (R^{i}_{\alpha}(\phi)\varepsilon^{\alpha})(x)\nonumber\\
&=& \sum_{\alpha}\int \,dy \,R^{i}_{\alpha}(x,y)\,
\varepsilon^{\alpha}(y).
\end{eqnarray}
The $\varepsilon^\alpha(x)$ are the infinitesimal gauge parameters and the 
$R^i_\alpha (\phi)$ the generators of the gauge transformations. When 
$\epsilon_\alpha=\epsilon(\varepsilon^\alpha)=0$ we have an ordinary 
symmetry, when $\epsilon_\alpha =1$ a supersymmetry. The Grassman parity 
of $R^i_\alpha$ is $\epsilon(R^i_\alpha)=\epsilon_i+
\epsilon_\alpha\,\,\, ({\rm mod\,2}).$  When
the gauge generators are independent the theory is said to be {\em irreducible}, 
otherwise it is reducible.  For our purposes it will be sufficient to 
consider the irreducible case.  

A subscript index after a comma denotes the right derivative with respect to 
the corresponding field, and in general when a derivative is indicated it is 
to be understood as a right derivative unless specifically noted to be 
otherwise. The field equations may then be written as
\begin{equation}
\mc{S}_{0,i} = 0.
\end{equation}
The classical solutions $\phi_0$ are determined by $\mc{S}_{0,i}|_{\phi_0}=0.$
The Noether identities are
\begin{equation}
\mc{S}_{0,i}\,R^i_\alpha=0.
\end{equation}
The general solution to the Noether identity $\mc{S}_{0,i}\,\lambda^i=0$ is
\begin{equation}
\lambda^i=R^i_\alpha\,T^\alpha + \mc{S}_{0,j}\,E^{ji}.
\label{sol}
\end{equation}

The commutator of two gauge transformations is
\begin{equation}
[\delta_{1},\delta_{2}]\phi^{i}=(R^{i}_{\alpha,j}R^{j}_{\beta}
-(-1)^{\epsilon_{\alpha}\epsilon_{\beta}}R^{i}_{\beta,j}R^{j}_{\alpha})
\varepsilon^{\beta}_{1}\varepsilon^{\alpha}_{2}.
\label{two}
\end{equation}
Since this commutator is a symmetry of the action it satisfies the
Noether identity
\begin{equation}
\mc{S}_{0,i}(R^{i}_{\alpha,j}R^{j}_{\beta}
-(-1)^{\epsilon_{\alpha}\epsilon_{\beta}}R^{i}_{\beta,j}R^{j}_{\alpha})=0,
\label{ni}
\end{equation}
 which by Eq. (\ref{sol}) implies that
\begin{equation}
R^{i}_{\alpha,j}R^{j}_{\beta}-(-1)^{\epsilon_{\alpha}\epsilon_{\beta}}
R^{i}_{\beta,j}R^{j}_{\alpha} = R^{i}_{\gamma}T^{\gamma}_{\alpha\beta}
-\mc{S}_{0,j}E^{ji}_{\alpha\beta}.
\label{TE}
\end{equation}
Eqs. (\ref{two}) and (\ref{TE}) lead to the following condition:
\begin{equation}[\delta_{1},\delta_{2}]\phi^{i}=(R^{i}_{\gamma}T^{\gamma}_{\alpha\beta}-
\mc{S}_{0,j}E^{ji}_{\alpha\beta})\varepsilon^{\beta}_{1}\varepsilon^
{\alpha}_{2}.
\label{Lie}
\end{equation}
The tensors $T^\gamma_{\alpha\beta}$ are called the structure constants of
the gauge algebra, although they depend in general on the fields of the 
theory.  When $E^{ij}_{\alpha\beta}=0$ the gauge algebra is said to be
{\em closed}, otherwise it is {\em open}. Eq. (\ref{Lie}) defines a 
{\em Lie algebra} if the algebra is closed and the $T^\gamma_{\alpha\beta}$ are 
independent of the fields.

The gauge tensors have the following graded symmetry properties:
\begin{eqnarray}
E^{ij}_{\alpha\beta} &=& -(-1)^{\epsilon_i\epsilon_j}
 E^{ji}_{\alpha\beta} =-(-1)^{\epsilon_\alpha\epsilon_\beta}
E^{ij}_{\beta\alpha},\\
T^\gamma_{\alpha\beta} &=& -(-1)^{\epsilon_\alpha\epsilon_\beta}
T^\gamma_{\beta\alpha}.
\end{eqnarray}
The Grassman parities are
\begin{equation}
\epsilon(T^\gamma_{\alpha\beta})=\epsilon_\alpha+\epsilon_\beta+
 \epsilon_\gamma\,\,\,({\rm mod\,2})
\end{equation}
and
\begin{equation}
\epsilon(E^{ij}_{\alpha\beta})=\epsilon_i+\epsilon_j+
 \epsilon_\alpha + \epsilon_\beta\,\,\,({\rm mod\,2}).
\end{equation}

Various restrictions are imposed by the Jacobi identity
\begin{equation}\sum_{cyclic(123)}[\delta_{1},[\delta_{2},\delta_{3}]]=0.
\end{equation}
These restrictions are
\begin{equation}
\sum_{cyclic(123)}(R^{i}_{\delta}A^{\delta}_{\alpha\beta\gamma}-
\mc{S}_{0,j}B^{ji}_{\alpha\beta\gamma})\varepsilon^{\gamma}_{1}
\varepsilon^{\beta}_{2}\varepsilon^{\alpha}_{3}=0,
\end{equation}
where
\[3A^{\delta}_{\alpha\beta\gamma} \equiv (T^{\delta}_{\alpha\beta ,k}
R^{k}_{\gamma}-T^{\delta}_{\alpha\eta}T^{\eta}_{\beta\gamma})\]
\begin{equation}
 +(-1)^{\epsilon_{\alpha}(\epsilon_{\beta}+\epsilon_{\gamma})}
(T^{\delta}_{\beta\gamma ,k}R^{k}_{\alpha}-T^{\delta}_{\beta\eta}
T^{\eta}_{\gamma\alpha})+
(-1)^{\epsilon_{\gamma}(\epsilon_{\alpha}+\epsilon_{\beta})}
(T^{\delta}_{\gamma\alpha ,k}R^{k}_{\beta}-T^{\delta}_{\gamma\eta}
 T^{\eta}_{\alpha\beta})
\end{equation}
and
\[3B^{ji}_{\alpha\beta\gamma}\equiv(E^{ji}_{\alpha\beta ,k}
R^{k}_{\gamma}-E^{ji}_{\alpha\delta}T^{\delta}_{\beta\gamma}
-(-1)^{\epsilon_{i}\epsilon_{\alpha}}R^{j}_{\alpha ,k}
E^{ki}_{\beta\gamma}+(-1)^{\epsilon_{j}(\epsilon_{i}+\epsilon_{\alpha})}
R^{i}_{\alpha ,k}E^{kj}_{\beta\gamma})\]
\[
+(-1)^{\epsilon_{\alpha}(\epsilon_{\beta}+\epsilon_{\gamma})}
(\alpha\rightarrow\beta,\;,\; \beta \rightarrow\gamma, \:\; 
\gamma\rightarrow\alpha)\]
\begin{equation}
+(-1)^{\epsilon_{\gamma}(\epsilon_{\alpha}+\epsilon_{\beta})}
(\alpha\rightarrow\gamma,\;\; \beta \rightarrow\alpha,\;\; 
\gamma\rightarrow\beta).
\end{equation}

As in the familiar Faddeev-Popov procedure it is useful to introduce ghost 
fields $C^{\alpha}$ with opposite Grassman parities to the gauge parameters 
$\varepsilon^\alpha$;
\begin{equation}
\epsilon(C^{\alpha})=\epsilon_{\alpha}+1,
\end{equation} 
and to replace the gauge parameters by ghost fields. One must then modify
the graded symmetry properties of the gauge structure tensors according to
\begin{equation}
T_{\alpha_1\alpha_2\alpha_3\alpha_4\ldots}\rightarrow (-1)^{\epsilon_{\alpha_2}+
\epsilon_{\alpha_4}+\ldots}\,\,T_{\alpha_1\alpha_2\alpha_3\alpha_4\ldots}
\;\;.
\end{equation}
The Noether identities then take the form
\begin{equation}
\mc{S}_{0,i}R^{i}_{\alpha}C^{\alpha}=0,
\label{noet}
\end{equation}
and the structure relations (\ref{TE}) become
\begin{equation}
(2R^{i}_{\alpha ,j}R^{j}_{\beta}-R^{i}_{\gamma}T^{\gamma}_{\alpha\beta}+
\mc{S}_{0,j}E^{ji}_{\alpha\beta})C^{\beta}C^{\alpha}=0.
\label{ekr}
\end{equation}

\subsection{Introducing the Antifields}

We incorporate the ghost fields into the field set 
$\Phi^A=\{\phi^i,C^\alpha\}$, where $i=1,\ldots,n$ and $\alpha=1,\ldots,m$. 
Clearly $A=1,\ldots,N$, where $N=n+m$. One then further increases the set by 
introducing an antifield $\Phi^{\star}_{A}$ for each field $\Phi^A$. 
The Grassman parity of the antifields is
\begin{equation}
\epsilon(\Phi^{\star}_{A})=\epsilon(\Phi^{A})+1\;\;({\rm mod\;2}).
\end{equation}
We also assign to each field  a {\em ghost number}, with
\begin{eqnarray}
{\rm gh}[\phi^i]&=&0,\\
{\rm gh}[C^\alpha]&=&1,\\
{\rm gh}[\Phi^{\star}_{A}]&=&-{\rm gh}[\Phi^{A}]-1\,.
\end{eqnarray}
In the space of fields and antifields the {\em antibracket} is defined by
\begin{equation}
(X,Y)=\frac{\partial_{r}X}{\partial\Phi^{A}}\frac{\partial_{l}Y}
{\partial\Phi^{\star}_{A}}
-\frac{\partial_{r}X}{\partial\Phi^{\star}_{A}}\frac{\partial_{l}Y}
{\partial\Phi^{A}},
\end{equation}
where $\partial_r$ denotes the right, $\partial_l$ the left derivative.
The antibracket is graded antisymmetric;
\begin{equation}
(X,Y)=-(-1)^{(\epsilon_{X}+1)(\epsilon_{Y}+1)}(Y,X).
\end{equation}
It satisfies a graded Jacobi identity
\begin{equation}
((X,Y),Z)+(-1)^{(\epsilon_{X}+1)(\epsilon_{Y}+\epsilon_{Z})}((Y,Z),X)+
(-1)^{(\epsilon_{Z}+1)
(\epsilon_{X}+\epsilon_{Y})}((Z,X),Y)\!=\!0.
\end{equation}
It is a graded derivation
\begin{eqnarray}
(X,YZ)&=&(X,Y)Z+(-1)^{\epsilon_Y\epsilon_X}(X,Z)Y,\nonumber\\
(XY,Z)&=&X(Y,Z)+(-1)^{\epsilon_X\epsilon_Y}Y(X,Z).
\end{eqnarray}
It has ghost number
\begin{equation}
{\rm gh}[(X,Y)]={\rm gh}[X]+{\rm gh}[Y]+1
\end{equation}
and Grassman parity
\begin{equation}
\epsilon((X,Y))=\epsilon(X)+\epsilon(Y)+1\;\;({\rm mod\, 2}).
\end{equation}
For bosonic fields
\begin{equation}
(B,B)=2\;\frac{\partial B}{\partial\Phi^{A}}\frac{\partial B}{\partial\Phi^{\star}_{A}},
\end{equation}
for fermionic fields
\begin{equation}
(F,F)=0,
\end{equation}
and for any $X$
\begin{equation}
((X,X),X)=0.
\end{equation}

If we group the fields and antifields together into the set
\begin{equation}
z^{a}=\{\Phi^{A},\Phi^{\star}_{A}\},\;\;\;a=1,...,2N,
\end{equation}
then the antibracket is seen to define a symplectic structure on the space of 
fields and antifields;
\begin{equation}
(X,Y)=\frac{\partial X}{\partial z^{a}}\zeta^{ab}\frac{\partial Y}
{\partial z^{b}}
\end{equation}
with
\begin{equation}
\zeta^{ab}=\left(\begin{array}{cc}
0 & \delta^{A}_{B} \\
-\delta^{A}_{B} & 0 \\
\end{array} \right).
\end{equation}
The antifields can be thought of as conjugate variables to the fields, since
 \begin{equation}
(\Phi^{A},\Phi^{\star}_{B})=\delta^{A}_{B}.
\end{equation}

\subsection{The Classical Master Equation}

Let
$\mc{S}[\Phi^{A},\Phi^{\star}_{A}]$ be a functional of the fields and 
antifields with the dimensions of an action, vanishing ghost number and even
Grassman parity. The equation
\begin{equation}
(\mc{S},\mc{S})=2\;\frac{\partial \mc{S}}{\partial\Phi^{A}}\frac{\partial \mc{S}}
{\partial\Phi^{\star}_{A}}=0
\end{equation}
is the {\em classical master equation}.  Solutions of the classical master 
equation with suitable boundary conditions turn out to be generating 
functionals for the gauge structures of the gauge theory.  $\mc{S}$
is also the starting point for the quantization of the theory.

One denotes by $\Sigma$ the subspace of stationary points of the action in the 
space of fields and antifields:
\begin{equation}
\Sigma = \left\{ z^a \left| \frac{\partial\mc{S}}{\partial z^a} =0 
\right.\right\}.
\end{equation}
Given a classical solution $\phi_0$ of $\mc{S}_0$ one possible stationary point
is
\begin{equation}
\phi^i=\phi^i_0,\,\,\,\,C^\alpha=0,\,\,\,\, \Phi^{\star}_{A}=0.
\end{equation}

An action $\mc{S}$ which satisfies the classical master equation has its own 
set of invariances:
\begin{equation}
\frac{\partial\mc{S}}{\partial z^{a}}\,\,R^{a}_{b}=0,
\label{inv}
\end{equation}
with
\begin{equation}
R^{a}_{b}=\zeta^{ac}\,\,\frac{\partial_{l}\partial_{r}\mc{S}}{\partial z^{c}\partial 
z^{b}}\;.
\end{equation}
This equation implies
\begin{equation}
R^c_a\,R^a_b\,|_\Sigma =0.
\end{equation}
We see that $R^a_b$ is nilpotent on-shell. A nilpotent $2N\times 2N$
matrix has rank less than or equal to $N$.  Let $r$ be the rank of the hessian
of $\mc{S}$ at the stationary point:
\begin{equation}
r={\rm rank}\left.\frac{\partial_l\partial_r\mc{S}}{
\partial z^a\partial z^b}\right|_\Sigma.
\end{equation}
We then have $r \leq N$. The relevant solutions of the classical master 
equation are those for which $r=N$. In this case the number of independent 
gauge invariances of the type in Eq. (\ref{inv}) equals the number of 
antifields. When at a later stage the gauge is fixed the non-physical 
antifields are eliminated.

To ensure the correct classical limit the proper solution must contain the 
classical action $\mc{S}_0$ in the sense that
\begin{equation}
\mc{S}\left.[\Phi^{A},\Phi^{\star}_A]\right|_{{\Phi^{\star}_{A}=0}}=
\mc{S}_{0}[\phi^{i}].
\end{equation}
The action $\mc{S}[\Phi^{A},\Phi^{\star}_{A}]$ can be expanded in a series
in the antifields, while maintaining vanishing ghost number
and even Grassman parity: 
\begin{equation}
\mc{S}[\Phi,\Phi^{\star}]=\mc{S}_{0}+\phi^{\star}_{i}R^{i}_{\alpha}C^{\alpha}
+C^{\star}_{\alpha}\frac{1}{2}T^{\alpha}_{\beta\gamma}(-1)^{\epsilon_{\beta}}
C^{\gamma}C^{\beta}+\phi^{\star}_{i}\phi^{\star}_{j}(-1)^{\epsilon_{i}}\frac{1}{4}E^{ji}_
{\alpha\beta}(-1)^{\epsilon_{\alpha}}C^{\beta}C^{\alpha}+\ldots .
\end{equation}
When this is inserted into the classical master equation one finds that
this equation implies the gauge structure of the classical theory 
(see e.g. Eq. (\ref{k5}) below).

\subsection{Gauge-fixing and Quantization}

Eq. (\ref{inv}) shows that the action $\mc{S}$ still possesses gauge 
invariances, and hence is not yet suitable for quantization via the path 
integral approach: a gauge-fixing procedure is necessary.  In the 
Batalin-Vilkovisky approach the gauge is fixed, and the antifields 
eliminated, by use of a gauge-fixing fermion $\Psi$ which has Grassman parity 
$\epsilon(\Psi)=1$ and ${\rm gh}[\Psi]=-1$. It is a functional of the 
fields $\Phi^A$ only; its relation to the antifields is
\begin{equation}
\Phi^\star_A=\frac{\partial \Psi}{\partial \Phi^A}.
\label{con}
\end{equation}
We define a surface in functional space
\begin{equation}
\Sigma_\Psi=\left\{ \left(\Phi^A,\Phi_A^\star \right)\left| \Phi_A^\star =
\frac{\partial \Psi}{\partial \Phi^A} \right. \right\},
\end{equation}
so that for any functional $X[\Phi,\Phi^\star]$
\begin{equation}
X\left|_{\Sigma_\Psi}\right. = X \left[ \Phi, \frac{\partial \Psi}
{\partial \Phi} \right].
\end{equation}

To construct a gauge-fixing fermion $\Psi$ of ghost number $-1$ one must
again introduce additional fields.  The simplest choice utilizes a trivial
pair $\bar{C_\alpha},\bar{\pi}_{\alpha}$ with
\begin{eqnarray}
\epsilon(\bar{C_\alpha}) &=& \epsilon_\alpha +1,\,\,\,
\epsilon(\bar{\pi}_{\alpha})=\epsilon_\alpha,\\
{\rm gh}[\bar{C_\alpha}] &=& -1,\;\;\;\;\;\;{\rm gh}[\bar{\pi}_{\alpha}]=0.
\end{eqnarray}
The fields $\bar{C_\alpha}$ are the Faddeev-Popov antighosts.
Along with these fields we include the corresponding antifields
$\bar{C}^{\star\alpha},\;\;\bar{\pi}^{\star\alpha}$.
Adding the term $\bar{\pi}_\alpha\bar{C}^{\star\alpha}$ to the action
$\mc{S}$ does not spoil its properties as a proper solution to the
classical master equation, and we get the non-minimal action
\begin{equation}
\mc{S}^{non}=\mc{S}+\bar{\pi}_\alpha\bar{C}^{\star\alpha}.
\end{equation}
The simplest possibility for $\Psi$ is
\begin{equation}
\Psi=\bar{C}_\alpha \chi^\alpha (\phi),
\label{psi}
\end{equation}
where $\chi^\alpha$ are the gauge-fixing conditions for the fields
$\phi$.
We denote the gauge-fixed action by
\begin{equation}
\mc{S}_\Psi = \mc{S}^{non}\left|_{\Sigma_\Psi} \right. .
\end{equation}

Quantization is performed using the path integral to calculate a 
correlation fuction $X$, with the constraint (\ref{con}) implemented by a
$\delta$-function:
\begin{equation}
I_\Psi (X) = \int \mc{D}\Phi \mc{D}\Phi^\star \delta\left(
\Phi_A^\star - \frac{\partial \Psi}{\partial \Phi^A} \right) 
\exp \left( \frac{i}{\hbar} W[\Phi,\Phi^\star] \right)
\,X[\Phi,\Phi^\star].
\label{int}
\end{equation}
Here $W$ is the quantum action, which reduces to $\mc{S}$ in the limit
$\hbar \rightarrow 0$.  An admissible $\Psi$ leads to well-defined
propagators when the path integral is expressed as a perturbation series
expansion.

The results of a calculation should be independent of the gauge-fixing.
Consider the integrand in Eq.(\ref{int}),
\begin{equation}
\mc{I}[\Phi,\Phi^\star]= \exp\left(\frac{i}{\hbar} W[\Phi,\Phi^\star]
\right)\, X[\Phi,\Phi^\star]\;.
\end{equation}
Under an infinitesimal change in $\Psi$
\begin{equation}
I_{\Psi+\delta\Psi}(X)-I_\Psi (X) \approx \int \mc{D}\Phi
\bigtriangleup\!\mc{I}\,\, \delta \Psi\,,
\end{equation}
where the Laplacian $\bigtriangleup$ is
\begin{equation}
\bigtriangleup = (-1)^{\epsilon_A +1} \frac{\partial}{\partial \Phi^A}
\frac{\partial}{\partial \Phi_A^\star}\;.
\end{equation}
Obviously the integral $I_\Psi (X)$ is independent of $\Psi$ if
$\bigtriangleup \mc{I} =0$. For $X=1$ one gets the requirement
\begin{equation}
\bigtriangleup \exp \left( \frac{i}{\hbar} W \right) =
\exp \left( \frac{i}{\hbar} W \right) 
\left( \frac{i}{\hbar} \bigtriangleup W -\frac{1}{2 \hbar^2} (W,W) \right)
=0.
\end{equation}
The formula
\begin{equation}
\frac{1}{2} (W,W)= i \hbar \bigtriangleup\!W
\label{qme}
\end{equation}
is the {\em quantum master equation}. A gauge-invariant correlation function
satisfies
\begin{equation}
(X,W)= i \hbar \bigtriangleup\!X.
\end{equation}

The terms of higher order in $\hbar$ by which the quantum action $W$
may differ from the solution of the classical master equation $\mc{S}$ 
correspond to the Counterterms of the renormalizable 
gauge theory if
\begin{equation}
\bigtriangleup \mc{S} = 0.
\end{equation}
One must of course use a regularization scheme which respects the symmetries
of the theory. For $W=\mc{S}+ O(\hbar)$ the quantum master equation
(\ref{qme}) reduces in this case to the classical master equation 
$(\mc{S},\mc{S})=0$. Hence, up to possible counter terms, one may simply choose
$W=\mc{S}$. This is the case for the systems we are
considering in this paper. 

To implement the gauge-fixing we use for the action $W=\mc{S}^{non}$.
For the path integral $Z=I_\Psi(X=1)$ we perform the integration over 
the antifields in Eq.(\ref{int}) by using the $\delta$-function. The result is
\begin{equation}
Z=\int \mc{D}\Phi \exp \left(\frac{i}{\hbar} \mc{S}_\Psi\right).
\label{PI}
\end{equation}

\section{The Path Integral for the Poisson-Sigma Model}
\subsection{The Classical Theory}

A {\em Poisson manifold} $N$ is a smooth manifold equipped with a Poisson
structure $P\in \Lambda^2TN$  \cite{VA}. In local coordinates $X^i$ on $N$
\begin{equation}
P=\frac{1}{2}\;P^{ij}(X)\;\partial_i\wedge \partial_j,
\end{equation}
and $P^{ij}$ satisfies the condition
\begin{equation}
P^{i[j}P^{lk]}{}_{,i}=0,
\label{SN}
\end{equation}
which reflects the vanishing of the corresponding Schouten-Nijenhuis bracket for $P$ with 
itself. Here the bracketed indices denote an antisymmetric sum. In the notation of Poisson 
brackets
\begin{equation}
\{f(X),g(X)\}=P^{ij}f_{,i}g_{,j}
\end{equation}
and the Jacobi identity follows from Eq. (\ref{SN}):
\begin{equation}
\{f,\{g,h\}\}+{\rm cyclic}=0.
\end{equation}
The Poisson bracket satisfies the Leibniz derivation rule:
\begin{equation}
\{h,fg\}=\{h,f\}g+f\{h,g\}\;.
\end{equation}
$P$ is in general degenerate, in which case it does not induce
a symplectic structure on $N$, and the map $T^\star N
\rightarrow TN$ induced by P, which maps a one-form $\alpha_i dX^i$ on $N$
to the vector field $\alpha_i P^{ij}\partial_j$, is not surjective. However,
as a consequence of the Jacobi identity, the image of this map forms an
involutive system of vector fields. It then turns out that the characteristic distribution 
$S(N)$ of the Poisson manifold $N$ is completely integrable and the Poisson structure 
$P$ induces symplectic structures on the leaves $S$, i.e. a nondegenerate symplectic 
structure $\Omega_S$ on $S$.

Indeed, the splitting theorem of Weinstein \cite{WE} states that for a regular Poisson 
manifold, i.e. the Poisson tensor has constant rank, there exist so-called Casimir-Darboux coordinates on the Poisson manifold $N$.
For $P$ degenerate there are functions $f$ on $N$ whose
Hamiltonian vector fields $X_f=f_{,i}P^{ij}\partial_j$ vanish. These
functions are called Casimir functions. Let $\{C^I\}$ be a maximal set of
independent Casimir functions. Then $C^I (X)={\rm const.} = C^I (X_0)$
defines a level surface through $X_0$ whose connected components may be 
identified with the symplectic leaves.
According to Darboux's theorem there are local coordinates 
$X^\alpha$ on $S$ such that the symplectic form $\Omega_S$ is given by
\begin{equation}
\Omega_S=dX^1\wedge dX^2 +dX^3\wedge dX^4 +\ldots.
\end{equation}
Together with the Casimir functions we then have a coordinate system
$\{X^I,X^\alpha\}$ on $N$ with $P^{IJ}=P^{I\alpha}=0$ and $P^{\alpha\beta}
= {\rm constant}$.

We now consider a field theory on a two dimensional world sheet $M$ without boundary 
\cite{SS1}.
The theory involves a set $X^i$ of bosonic scalar fields, which can be interpreted as 
a set of maps $X^i:M\rightarrow N$. In addition one has a 1-form $A$ on the world sheet $M$ 
which takes values in $T^{\ast}N$, i.e. a 1-form on $M$ which is simultaneously the pullback 
of a section of $T^{\ast}N$ by $X(x)$, where $\{x\}$ are coordinates on M. This field 
$A=A_{\mu i}dx^{\mu}\wedge dX^{i}$ reduces in the case of a linear Poisson structure, 
which leads to the Yang-Mills theory, to an ordinary gauge field.
In these coordinates the action of the semi-topological Poisson-Sigma model is
\begin{equation}
\mc{S}_{0}[X,A]=\int\limits_{M}\:\mu\;\left[\epsilon^{\mu\nu}(A_{\mu i}\partial_{\nu}X^{i}+
\frac{1}{2}\;P^{ij}(X)A_{\mu i}A_{\nu j})+C(X)\right],
\label{CA}
\end{equation}
where $\mu$ is the volume form on $M$ and $C(X)$ is a Casimir function.

The gauge transformations are
\begin{equation}
\delta X^{i}=P^{ij}(X)\varepsilon_{j}\;,\;\;\;\delta A_{\mu i}=
D_{\mu i}^{j}\varepsilon_{j},
\end{equation}
where 
$D_{\mu i}^{j}=\partial_\mu \delta_i^j+P^{kj}{}_{,i}
A_{\mu k}$.
The equations of motion are
\begin{equation}\epsilon^{\mu\nu}D_{\mu i}^{j}A_{\nu j}+\frac{\partial C(X)}
{\partial X^{i}}=0
\end{equation}
and
\begin{equation}
\epsilon^{\mu\nu} \left(\partial_\nu X^i+P^{ij}A_{\nu j}\right)=\epsilon^{\mu\nu}
D_{\nu}X^i=0.
\end{equation}
The gauge algebra is given by
\begin{equation}
 [\delta(\varepsilon_{1}),\delta(\varepsilon_{2})]X^{i}=P^{ji}(P^{mn}{}_{,j}\;
\varepsilon_{1n}\varepsilon_{2m})\;,
\end{equation}
\begin{equation}
 [\delta(\varepsilon_{1}),\delta(\varepsilon_{2})]A_{\mu i}=D_{\mu i}^{j}(P^{mn}{}_{,j}
\varepsilon_{1n}\varepsilon_{2m})-(\epsilon^{\nu\rho}D_{\rho}X^{j})
\epsilon_{\nu\mu}P^{mn}{}_{,ji}\;\varepsilon_{1n}\varepsilon_{2m}
\end{equation}

In the notation of Section \ref{antifield} the generators of the gauge 
transformations $R$ are here $P^{ij}$ and $D^j_{\mu i}$. The gauge tensors
$T$ and $E$ are $P^{ij}{}_{,k}$ and $\epsilon_{\nu \mu} P^{mn}{}_{,ji}$.
The higher order gauge tensors $A$ and $B$ vanish.

We denote the ghost fields again by $C^i$. The Noether identities are then
\begin{equation}
\int\limits_{M}\,\mu\,\left((\epsilon^{\mu\nu}D_{\mu i}^{j}A_{\nu j}+\frac{\partial C(X)}
{X^i})P^{ki}+(\epsilon^{\mu\nu}D_{\nu}X^{i})D_{\mu i}^{k}\right)C_{k}=0.
\label{NI}
\end{equation}
Considering the commutator of two gauge transformations leads to
(see Eqs. (\ref{two}-\ref{TE}))
\begin{equation}
\int\limits_{M}\;\mu\;(2P^{mi}{}_{,j}P^{nj}-P^{ji}P^{mn}{}_{,j})C_{m}C_{n}=0
\label{2G}
\end{equation}
\begin{equation}
\int\limits_{M}\;\mu\;(2(P^{jk}{}_{,i}D_{\mu j}^{l}+P^{mk}{}_{,ij}A_{\mu m}P^{jl})-
D_{\mu i}^{m}P^{kl}{}_{,m}+(\epsilon^{\rho\nu}D_{\rho}X^{i})\epsilon_{\mu\nu}
P^{kl}{}_{,ji})C_{l}C_{k}=0.
\label{3G}
\end{equation}
The Jacobi identity is
\begin{equation}
P^{ij}{}_{,m}P^{mk}C_{i}C_{j}C_{k}=0.
\end{equation}
We shall later need the first derivative of the Jacobi identity:
\begin{equation}
(P^{ij}{}_{,mn}P^{mk}+P^{ij}{}_{,m}P^{mk}{}_{,n})C_{i}C_{j}C_{k}=0,
\label{FD}
\end{equation}
as well as the second derivative
\begin{equation}
(P^{ij}{}_{,mnp}P^{mk}+P^{ij}{}_{,mn}P^{mk}{}_{,p}+P^{ij}{}_{,mp}P^{mk}{}_{,n}
+P^{ij}{}_{,m}P^{mk}{}_{,np})C_{i}C_{j}C_{k}=0\;.
\label{SD}
\end{equation}
\bigskip\\
{\large\bf The Antifields of the Poisson-Sigma Model}

The fields and antifields of the model are
\begin{equation}
\Phi^{A}=\{A^{\mu i},X^{i},C_{i}\}\;\;{\rm and}\;\;\Phi^{\star}_{A}=
\{A^{\mu i\star},X^{\star}_{i},C^{i\star}\}\;.
\end{equation}
The extended action is
\begin{equation}
\begin{split}
\mc{S} &= \int\limits_{M}\;\mu\;\bigg[\epsilon^{\mu\nu}(A_{\mu i}\partial_{\nu}X^{i}+P^{ij}(X)
A_{\mu i}A_{\nu j})+C(X)+A^{\mu i\star}D_{\mu i}^{j}C_{j}+X^{\star}_{i}
P^{ji}(X)C_{j}\\
& \phantom{\int\limits_{M}+\,+\bigg[} +\frac{1}{2}C^{i\star}P^{jk}{}_{,i}(X)C_{j}C_{k}+
\frac{1}{4}A^{\mu i\star}A^{\nu j\star}\epsilon_{\mu\nu}P^{kl}{}_{,ij}(X)C_{k}C_{l}\bigg].
\label{ea}
\end{split}
\end{equation}
The classical master equation is
\begin{gather}
 (\mc{S},\mc{S}) =\int\limits_{M}\;\mu\;\bigg[(\epsilon^{\nu\rho}(D_{\nu}X^{m}
)D^{j}_{\rho m}
+(\epsilon^{\mu\nu}D^{i}_{\mu m}A_{\nu i}+\frac{\partial C(X)}{X^i})P^{jm})C_{j}\notag \\
 -(X^{\star}_{i}P^{ij}{}_{,m}P^{km}-X^{\star}_{i}P^{im}\frac{1}{2}P^{jk}{}_{,m})C_{j}C_{k}
\notag \\
  +\epsilon^{\mu\rho}(D_{\mu}X^{m})A^{\nu j\star}\epsilon_{\rho\nu}\frac{1}{2}
P^{kl}{}_{,mj}C_{k}C_{l}-A^{\rho i\star}P^{mk}{}_{,i}C_{k}D^{j}_{\rho m}C_{j}\notag \\
 -A^{\mu i\star}P^{jk}{}_{,im}A_{\mu j}C_{k}P^{mn}C{n}-(D^{m}_{\mu i}A^{\mu i\star})
\frac{1}{2}P^{jk}{}_{,m}C_{j}C_{k} \notag\\
 +\frac{1}{2}C^{i\star}P^{jk}{}_{,im}C_{j}C_{k}P^{lm}C_{l}+C^{i\star}P^{mk}{}_{,i}C_{k}
\frac{1}{2}P^{jl}{}_{,m}C_{j}C_{l}\notag\\
 +A^{\mu i\star}A^{\nu j\star}\epsilon_{\mu\nu}(\frac{1}{4}P^{kl}{}_{,ijm}C_{k}C_{l}
P^{mn}C_{n}+\frac{1}{4}P^{ml}{}_{,ij}C_{l}P^{kl}{}_{,m}C_{k}C_{l}-\frac{1}{2}P^{mn}{}_{,i}C_{n}
P^{kl}{}_{,mj}C_{k}C_{l})\bigg]=0\;.\notag\\
{}
\label{k5}
\end{gather}
Eqs. (\ref{NI})-(\ref{SD}) ensure that 
the extended action (\ref{ea}) is a solution of the classical master equation (\ref{k5}).
\bigskip\\
{\large\bf Gauge-fixing}

We shall use gauge-fixing conditions of the form $\chi_i(A,X)$, so that the
gauge fermion (\ref{psi}) becomes $\Psi=\bar{C}^i \chi_i(A,X)$. 
The antifields are then fixed to be
\begin{equation}
\begin{split}
A^\star_{\mu i} &= \bar{C}_{j}\frac{\partial \chi_j(A,X)}{\partial A_{\mu i}},\\
X^\star_i &= \bar{C}_{j}\frac{\partial \chi_j(A,X)}{\partial X^{i}},\\
C^\star_i &= 0,\\
\bar{C}^\star_i &= \chi_i(A,X).\\
\end{split}
\end{equation}
The gauge-fixed action is
\begin{equation}
\begin{split}
\mc{S}_\Psi &= \int\limits_{M}\,\mu\;\bigg[\epsilon^{\mu\nu}(A_{\mu i}\partial_{\nu}X^{i}+
P^{ij}(X)A_{\mu i}A_{\nu j})+C(X)\\
& \phantom{\int\limits_{M}+\,+\bigg[}+\bar{C}^k\frac{\partial\chi_k(A,X)}
{\partial A_{\mu i}}D_{\mu i}^{j}C_{j}+\bar{C}^{k}\frac{\partial\chi_{k}(A,X)}
{\partial X^{i}}P^{ij}C_{j}\\
& \phantom{\int\limits_{M}+\,+\bigg[}+\frac{1}{4}\bar{C}^m\frac{\partial\chi_m(A,X)}{\partial 
A_{\mu i}}\bar{C}^n\frac{\partial\chi_n(A,X)}{\partial A_{\nu j}}\epsilon_{\mu\nu}
P^{kl}{}_{,ij}(X)C_{k}C_{l}\}+\bar{\pi}^i\chi_i(A,X)\bigg].
\end{split}
\end{equation}

We now consider different gauge conditions:

(i) First, the Landau gauge for the gauge potential $\chi_{i}=
\partial^{\mu}A_{\mu i}$, so that the gauge fermion becomes $\Psi=\bar{C}^{i}
\partial^{\mu}A_{\mu i}$. The antifields are fixed to be:
\begin{equation}
\begin{split}
A^{\star\mu i} &= \partial^{\mu}\bar{C}^{i},\\
X^{\star}_{i} &= C^{\star i}=0,\\
\bar{C}^{\star}_{i} &= \partial^{\mu}A_{\mu i}.\\
\end{split}
\end{equation}
For this gauge choice the gauge-fixed action is:
\begin{equation}
\begin{split}
\mc{S}_\Psi &= \int\limits_{M}\;\mu\bigg[\epsilon^{\mu\nu}(A_{\mu i}\partial_{\nu}
X^{i}+P^{ij}(X)A_{\mu i}A_{\nu j})+C(X)+\bar{C}^i\partial^{\mu}D_{\mu i}^{j}C_{j}\\
& \phantom{\int\limits_{M}+\,+\bigg[} +\frac{1}{4}(\partial^{\mu}\bar{C}^i)(\partial^{\nu}
\bar{C}^j)\epsilon_{\mu\nu}P^{kl}{}_{,ij}(X)C_{k}C_{l}-\bar{\pi}^i(\partial^{\mu}A_{\mu i})
\bigg].
\end{split}
\end{equation}
Translating this action into the notation of Cattaneo and Felder \cite{CF} 
one sees that it is exactly the expression they use to derive the 
pertubation series.  

(ii)
Now we consider the temporal gauge $\chi_i=A_{0i}$. In this case the gauge fermion 
is given by $\Psi=\bar{C}^{i}A_{0i}$. The antifields are fixed to:
\begin{equation}
\begin{split}
A^{\star 0i} &= \bar{C}^{i},\\
A^{\star 1i} &= 0,\\
X^{\star}_{i} &= C^{\star i}=0,\\
\bar{C}^{\star}_{i} &= A_{0i}.\\
\end{split}
\end{equation}
The gauged-fixed action is:
\begin{equation}
\mc{S}_\Psi = \int\limits_{M}\;\mu\;\bigg[\epsilon^{\mu\nu}(A_{\mu i}\partial_{\nu}
X^{i}+P^{ij}(X)A_{\mu i}A_{\nu j})+C(X)+\bar{C}^iD_{0 i}^{j}C_{j}-
\bar{\pi}^i(A_{0i})\bigg].
\end{equation}

(iii) Finally we consider the Schwinger-Fock gauge $\chi_i=x^{\mu}A_{\mu i}$. 
Then the antifields 
are fixed to be:
\begin{equation}
\begin{split}
A^{\star\mu i} &= x^{\mu}\bar{C}^{i},\\
X^{\star}_{i} &=  C^{\star i}=0,\\
\bar{C}^{\star}_{i} &= x^{\mu}A_{\mu i}.\\
\end{split}
\end{equation}
For this gauge choice the gauge-fixed action is:
\begin{equation}
\mc{S}_\Psi = \int\limits_{M}\;\mu\;\bigg[\epsilon^{\mu\nu}(A_{\mu i}
\partial_{\nu}X^{i}+P^{ij}(X)
A_{\mu i}A_{\nu j})+C(X)+\bar{C}^{i}x^{\mu}D_{\mu i}^{j}C_{j}
-\bar{\pi}^i(\partial^{\mu}A_{\mu i})\bigg].
\end{equation}

Notice that in the non-covariant gauges (ii) and (iii) the action simplifies,
in that the term which arose because of the 
non-closed gauge algebra vanishes.
\bigskip\\
{\large\bf Gauge fixing in Casimir-Darboux coordinates}

Important simplifications occur when we write the action 
in Casimir-Darboux coordinates $X^i\rightarrow\{X^I,X^\alpha\}$,
so we go through the gauge-fixing procedure again for these coordinates.
The extended action is
\begin{equation}
\begin{split}
\mc{S} &= \int\limits_{M}\;\mu\;\Bigg[\epsilon^{\mu\nu}(A_{\mu I}\partial_{\nu}X^{I}+
A_{\mu\alpha}\partial_{\nu}X^{\alpha}+P^{\alpha\beta}(X^{I}) A_{\mu\alpha}A_{\nu\beta})
+C(X^I)+A^{\mu I\star}\partial_{\mu}C_{I}\\
&  +A^{\mu\alpha\star}\partial_{\mu}C_{\alpha}+X^{\star}_{\alpha}P^{\beta\alpha}(X^{I})
C_{\beta}\bigg].\phantom{A_{\nu\beta})+C(X^I)+A^{\mu I\star}\partial_{\mu}C_{I}}
\end{split}
\end{equation}
This extended action still possesses gauge invariances, so one has to introduce a nonminimal 
sector. The non-minimal action is
\begin{equation}
\begin{split}
\mc{S}^{non} &= \int\limits_{M}\;\mu\;\Bigg[\epsilon^{\mu\nu}(A_{\mu I}\partial_{\nu}X^{I}
+A_{\mu\alpha}\partial_{\nu}X^{\alpha}+P^{\alpha\beta}(X^{I})A_{\mu\alpha}A_{\nu\beta})
+ C(X)\\
&   \phantom{\int\limits_{M}+\,+\Bigg[} +A^{\mu I\star}\partial_{\mu}C_{I}+A^{\mu\alpha\star}
\partial_{\mu}C_{\alpha}+
X^{\star}_{\alpha}P^{\beta\alpha}(X^{I})C_{\beta} -\bar{\pi}^{I}\bar{C}^{\star}_{I}-
\bar{\pi}^{\alpha}\bar{C}^{\star}_{\alpha}\bigg].
\end{split}
\end{equation} 
In these coordinates the gauge freedom of the maps $X^i:M\rightarrow N$ is reduced to 
the freedom of the maps $X^\alpha: M\rightarrow S$, where $S$ is a symplectic leaf of the 
Poisson manifold $N$. The gauge transformations $\delta_\varepsilon X^i=P^{ij}\varepsilon_j$ 
reduce to
\begin{equation}
\delta_\varepsilon X^\alpha=P^{\alpha \beta} \varepsilon_\beta,\;\;\;
\delta_\varepsilon X^I=0.
\end{equation}
After gauge fixing we need to consider only the homotopy classes $[X^\alpha]$. 

It is now possible to decompose the gauge condition into a part depending only 
on $A_{\mu I}$ and another part depending only on $X^{\alpha}$, so that the 
gauge-fixing of the gauge fields is implemented by gauge conditions 
of the form  $\chi_I(A_{I})$ and $\chi_\alpha(X^\alpha)$.
The gauge fermion may be written as
\begin{equation}
\Psi=\int\limits_M\;\mu\;\left[\bar{C}^I\chi_{I}(A_{I})+\bar{C}^\alpha\chi_{\alpha}(X^{\alpha})\right].
\end{equation}

The gauge conditions as expressed through the gauge fermion are
\begin{equation}
\begin{split}
A^{\mu I\star} &= \bar{C}^{J}\frac{\partial \chi_J (A_{I})}{\partial A_{\mu I}},\\
A^{\mu\alpha\star} &= 0,\\
X^{\star}_{\alpha} &= \bar{C}^{\beta}\frac{\partial\chi_{\beta}(X^{\alpha})}
{\partial X^{\alpha}},\\
C^{\star}_{i} &= 0,\\
\bar{C}^{\star}_{I} &=\chi_I(A_{I}),\\
\bar{C}^{\star}_{\alpha} &=\chi_\alpha(X^{\alpha})\,.\\
\end{split}
\end{equation}

The gauge-fixed action in Casimir-Darboux coordinates takes the form
\begin{equation}
\begin{split}
\mc{S}_{\psi} &= \int\limits_{M}\;\mu\;\Bigg[\epsilon^{\mu\nu}(A_{\mu I}\partial_{\nu}X^{I}+
A_{\mu\alpha}\partial_{\nu}X^{\alpha}+P^{\alpha\beta}A_{\mu\alpha}A_{\nu\beta}))
+C(X^I)\\
&  \phantom{\int\limits_{M}+\,+\Bigg[}+\bar{C}^{J}\frac{\partial\chi_J(A_{J})}
{\partial A_{\mu I}}\partial_\mu C_{I}+
\bar{C}^{\alpha}\frac{\partial\chi_{\alpha}(X^{\alpha})}{\partial X^{\beta}}
P^{\beta\gamma}C_{\gamma}-\bar{\pi}^{I}\chi_{I}(A_{I})-\bar{\pi}^{\alpha}\chi_{\alpha}
(X^{\alpha})\Bigg].
\end{split}
\end{equation}

\subsection{The Path Integral for the Poisson-Sigma Model}

Using Eq. (\ref{PI}) the path integral for the Poisson-Sigma model 
in Casimir-Darboux coordinates is
\begin{equation}
Z=\int_{\Sigma_\Psi} \mc{D}X^I\mc{D}X^\alpha\mc{D}A_{\mu I}
\mc{D}A_{\mu \alpha}\mc{D}C_I\mc{D}\bar{C}_I\mc{D}C_\alpha
\mc{D}\bar{C}_\alpha\mc{D}\bar{\pi}_I\mc{D}\bar{\pi}_\alpha
\exp \left(-\frac{1}{\hbar}\mc{S}_\Psi\right),
\end{equation}
where we have performed the usual Wick rotation $t=i\tau$, so that the exponent of the 
path integral is now real. When the model is integrable we expect to be able to carry out 
the functional integrations successively, in order to obtain a closed expression for the 
path integral.  We shall indeed be able to achieve this goal for the special case 
described in Section (4).

Integrating over the ghost and antighost fields yields the Faddeev-Popov determinants:
\begin{gather}
Z=\int_{\Sigma_\Psi}\!\!\mc{D}X^I\mc{D}X^\alpha\mc{D}A_{\mu I}
\mc{D}A_{\mu \alpha}\mc{D}\bar{\pi}_I\mc{D}\bar{\pi}_\alpha
{\rm det}\left(\frac{\partial\chi_{I}(A_{I})}
{\partial A_{\mu I}}\partial_{\mu} \right)_{\Omega^{0}(M)}\!\!\!
{\rm det}\left( \frac{\partial\chi_{\alpha}(X^{\alpha})}{\partial X^{\gamma}}
P^{\gamma\beta}(X^{I})\right)_{\Omega^{0}(M)}\notag\\
\exp\!\left(\!-\frac{1}{\hbar}\int\limits_{M}\!\mu\Bigg[\epsilon^{\mu\nu}(A_{\mu I}\partial_{\nu}
X^{I}+A_{\mu\alpha}\partial_{\nu}X^{\alpha}+P^{\alpha\beta}A_{\mu\alpha}
A_{\nu\beta}))+C(X^I)\!-\!\bar{\pi}^{I}\chi_I(A)\!-\!\bar{\pi}^{\alpha}
\chi_{\alpha}(X^{\alpha})\Bigg]\!\right),\notag\\
{}
\end{gather}
where the subscripts $\Omega^{k}(M)$ indicate that the determinant results from an 
integration over k-forms on $M$.
The integrations over $\bar{\pi}_I$ and $\bar{\pi}_\alpha$ yield $\delta$-
functions which implement the gauge conditions. 
\begin{gather}
Z = \int_{\Sigma_\Psi} \mc{D}X^I\mc{D}X^\alpha\mc{D}A_{\mu I}
\mc{D}A_{\mu \alpha}\;{\rm det}\left( \frac{\partial\chi_I(A_{I})}{\partial A_{\mu I}}
\partial_{\mu} \right)_{\Omega^{0}(M)}{\rm det}\left(\frac{\partial\chi_\alpha(X^{\alpha})}
{\partial X^{\gamma}}P^{\gamma\beta}(X^{I})\right)_{\Omega^{0}(M)}\notag\\
 \exp\left(-\frac{1}{\hbar}\int\limits_{M}\;\mu\;\Bigg[\epsilon^{\mu\nu}(A_{\mu I}
\partial_{\nu}X^{I}+A_{\mu\alpha}\partial_{\nu}X^{\alpha}+
P^{\alpha\beta}A_{\mu\alpha}A_{\nu\beta}))+C(X^I)\Bigg]\right),
\end{gather}
where from now on the integrations extend only over the degrees of freedom 
which respect the gauge-fixing conditions. 
The integration over $A_{\mu \alpha}$ is gaussian, it yields 
\begin{gather}
Z = \int_{\Sigma_\Psi}\mc{D}X^I\mc{D}X^\alpha\mc{D}A_{\mu I}\;{\rm det}\left(
\frac{\partial\chi_I(A^{I})}{\partial A_{\mu I}}\partial_{\mu}\right)_{\Omega^{0}(M)}
{\rm det}\left(\frac{\partial\chi_\alpha(X^{\alpha})}{\partial X^{\gamma}}
P^{\gamma\beta}(X^{I})\right)_{\Omega^{0}(M)}\notag\\
 {\rm det}^{-1/2}\left(P^{\alpha\beta}(X^{I})\right)_{\Omega^{1}(M)} 
\exp \left(-\frac{1}{\hbar}\int\limits_{M}\;\mu\;\Bigg[\epsilon^{\mu\nu}
( A_{\mu I}\partial_{\nu}X^{I}+\Omega_{\alpha \beta}\partial_\mu X^\alpha \partial_\nu 
X^\beta )+C(X^I)\Bigg]\right).\notag\\
{}
\end{gather}
Besides the term in the exponent the only dependence on $A_{\mu I}$ is in the relevant 
Faddeev-Popov determinant. If we choose a gauge condition linear in $A_{\mu I}$ this 
determinant becomes independent of the fields, and can be absorbed into a normalization 
factor. The integration over $A_{\mu I}$ then yields a $\delta$-function for 
$\partial_\nu X^I$. When this  $\delta$-function is implemented the fields $X^I$ become 
independent of the coordinates $\{x^\mu\}$ on $M$. Hence the Casimir functions are 
constants. The constant modes of the Casimir coordinates $X_0^I$ count the symplectic 
leaves. The path integral is now
\begin{gather}
Z = \int_{\Sigma_\Psi} dX_0^I\mc{D}X^{\alpha}\; 
{\rm det}\left( \frac{\partial\chi_\alpha(X^{\alpha})}{\partial X^\gamma}
P^{\gamma\beta}(X^{I}_0)\right)_{\Omega^{0}(M)}{\rm det}^{-1/2}\left(P^{\alpha\beta}(X^{I}_0)
\right)_{\Omega^{1}(M)}\notag\\
 \exp \left(-\frac{1}{\hbar}\int\limits_M\;\mu\; \Omega_{\alpha \beta}dX^\alpha dX^\beta\right)
\exp\left(-\int\limits_M \frac{1}{\hbar}\mu C(X_0^I)\right).
\end{gather}
The gauge-fixing of the fields $X^\alpha$ reduces the integral $\mc{D}X^\alpha$ to a sum 
over the homotopy classes:
\begin{gather}
Z = \int_{\Sigma_\Psi} dX_0^I \sum\limits_{[M\rightarrow S(X_0^I)]} 
{\rm det}\left( \frac{\partial\chi_\alpha(X)}{\partial X^{\gamma}}
P^{\gamma\beta}(X^{I}_0)\right)_{\Omega^{0}(M)}
{\rm det}^{-1/2}\left(P^{\alpha\beta}(X^{I}_0)\right)_{\Omega^{1}(M)}\notag\\
 \exp\left(-\frac{1}{\hbar}\int\limits_M\;\mu\; \Omega_{\alpha\beta}dX^\alpha dX^{\beta}\right)
\exp\left(-\frac{1}{\hbar}\int\limits_M \mu C(X_0^I)\right).
\end{gather}
Since the $C(X_0^I)$ are independent of the coordinates on $M$ the
last exponent simplifies to
\begin{equation}
\exp\left(-\frac{1}{\hbar}\int\limits_M \mu C(X_0^I)\right)=\exp(-\frac{1}{\hbar}A_M C(X_0^I)),
\end{equation}
where $A_M$ is the surface area of $M$. The form of the path integral  
then becomes
\begin{gather}
Z = \int_{\Sigma_\Psi} dX_0^I \sum\limits_{[M\rightarrow S(X_0^I)]} 
{\rm det}\left( \frac{\partial\chi_\alpha(X)}{\partial X^{\gamma}}
P^{\gamma\beta}(X^{I}_0)\right)_{\Omega^{0}(M)}
{\rm det}^{-1/2}\left(P^{\alpha\beta}(X^{I}_0)\right)_{\Omega^{1}(M)}\notag\\
\exp\left(-\frac{1}{\hbar}\int\limits_M\;\mu\;\Omega_{\alpha\beta}dX^\alpha dX^{\beta}\right)
\exp(-\frac{1}{\hbar}A_M C(X_0^I)).
\label{pf}
\end{gather}
Note that we have now arrived at an almost closed expression for the partition function for the Poisson-Sigma model, i.e. all the functional integrations have been performed.
\section{SU(2) Yang-Mills Theory}

To make further progress we consider the special case where the Poisson manifold 
$N=\mathbb{R}^3$, 
and the Poisson structure is linear: $P^{ij}=c^{ij}_k X^k$. The quadratic Casimir operator is 
$C(X)=\sum_i X^i X^i$. If we use this Casimir operator in the classical action (\ref{CA}) we 
may integrate out the $X^i$ fields to obtain the action for the two-dimensional Yang-Mills 
theory. If we omit the Casimir term in the action the same procedure yields the topological 
BF-theory.

Because of the Jacobi identity the structure constants $c^{ij}_{k}$  define a Lie Algebra 
structure on the dual space $\mc{G}$ of $N$. For this reason the linear Poisson structure is 
also called a Lie-Poisson structure on $N$. We are here interested in the case where the 
Lie algebra is three-dimensional, and the structure constants are those of the group $SU(2)$. 
The Poisson structure is degenerate and the symplectic leaves are two dimensional spheres 
characterized, in the Casimir-Darboux coordinates, by their radius $X^I_0$. Weinstein 
\cite{WE} has shown that the symplectic leaves of a linear Poisson structure are the 
co-adjoint orbits of the corresponding compact, connected Lie group G of $\mc{G}$. By a 
theorem of Kirillov these orbits can in turn be identified with the irreducible unitary 
representations of G \cite{KI1}. 

These considerations can be used to further reduce the expression for the
path integral. Consider the homotopy classes of the maps $X^{\alpha}:M\lra S(X^{I}_0)$.
The Hopf theorem tells us that the mappings $ f,g:M\lra S(X^{I}_0)$ are 
homotopic if and only if the degree of the mapping $f$ is the same as the degree of $g$. 
This means that the sum over the homotopy classes of the maps $[X^{\alpha}]$ can be expressed 
as a sum over the degrees $n={\rm deg}[X^{\alpha}]$:
\begin{equation}
\sum_{[X^{\alpha}]}\lra \sum_{n\in\mathbb{Z}}
\end{equation}
For  a map $f:X\lra Y$, where $X$ and $Y$ are k-dimensional oriented manifolds
and $\omega$ a k-form on $Y$, the degree of the mapping is given by 
\begin{equation}
 \int\limits_{X}\,f^{\ast}\omega={\rm deg}[f]\int\limits_{Y}\omega\;.
\end{equation}
Using this formula yields:
\begin{equation}
 \int\limits_{M}\;\mu\;\Omega_{\alpha\beta}{\rm d}X^{\alpha}{\rm d}X^{\beta}=n\int
\limits_{S}\Omega_{S}(X^I_0)\;,
\end{equation}
where $\Omega_{S}(X^I_0)$ is the symplectic form on the corresponding leaf $S$.
This gives for the partition function of Eq. (\ref{pf})
\begin{gather}
 Z = \int_{\Sigma_\Psi}{\rm d}X^{I}_{0}\sum_{n\in\mathbb{Z}}
{\rm det}\left( \frac{\partial\chi_\alpha(X)}{\partial X^{\gamma}}
P^{\gamma\beta}(X^{I}_0)\right)_{\Omega^{0}(M)}{\rm det}^{-1/2}\left(P^{\alpha\beta}(X^{I}_0)
\right)_{\Omega^{1}(M)}\notag\\
 \times\;\exp\{-\,n\int\limits_{S}\Omega_{S}(X^I_0)\}\exp(-\frac{1}{\hbar}A_{M}C(X^{I}_{0})).
\end{gather}
The sum over $n$ yields a periodic $\delta$-function:
\begin{gather}
Z  = \int_{\Sigma_\Psi}{\rm d}X^{I}_{0}\sum_{n\in\mathbb{Z}}{\rm det}\left( 
\frac{\partial\chi_\alpha(X)}{\partial X^{\gamma}}P^{\gamma\beta}(X^{I}_0)\right)_{\Omega^{0}
(M)}{\rm det}^{-1/2}\left(P^{\alpha\beta}(X^{I}_0)\right)_{\Omega^{1}(M)}\notag\\
\times \delta\left(\int\limits_{S}\Omega_{S}(X^I_0)-n\right)\exp(-\frac{1}{\hbar}A_{M}
C(X^{I}_{0})).
\label{per}
\end{gather}
The $\delta$-function says that the symplectic leaves must be integral. By the
identification of the leaves with the co-adjoint orbits, the orbits must also be integral. 
The fact that the orbits are integral reduces the number of the co-adjoint orbits to a 
countable set, which we label by $\mc{O}(\Omega)$.

We now consider the two determinants in the path integral. 
We choose the ``unitary gauge'' $\chi_\alpha(X^\alpha)=X^\alpha$,
so that $\partial \chi_\alpha(X) / \partial X^\gamma = \delta_\gamma^\alpha$,
and the two determinants have the same form.
The restiction of the scalar fields to the Casimir-Darboux coordinates $X^I$corresponds 
to the restriction of the scalar fields to the invariant Cartan subalgebra
considered by Blau and Thompson in \cite{BT2}, so we may adopt their 
argumentation concerning the powers to which the determinants occur for a manifold with
Euler characteristic $\chi(M)$. The result is a factor
\begin{equation}
{\rm det}(P^{\alpha\beta}(X^I_0))^{\chi(M)}.
\end{equation}

The determinant of a mapping equals the volume of the image of that mapping,
hence the determinant ${\rm det}(P^{\alpha\beta}(X^I_0))$ corresponds to  the 
symplectic volume of the leaf, which we denote by ${\rm Vol}(\Omega_S(X^I_0))$. 
The path integral then takes the form:
\begin{equation}
Z=\int_{\Sigma}{\rm d}X^{I}_{0}\sum_{n\in\mathbb{Z}}
{\rm Vol}(\Omega_{S}(X^{I}_{0}))^{\chi(M)}\;\delta\left(\int\limits_{S}\Omega_{S}(X^I_0)-n\right)\exp(-\frac{1}{\hbar}A_{M}C(X^{I}_{0})).
\end{equation}
Implementing the $\delta$-function by integrating over $X^{I}_{0}$
the sum over the mapping degrees becomes a sum over the set $\mc{O}(\Omega)$ of the integral 
orbits:
\begin{equation}
Z=\sum_{\mc{O}(\Omega)}{\rm Vol}(\Omega_{S}(X^{I}_{0}))^{\chi(M)}
\exp(-\frac{1}{\hbar}A_{M}C(X^{I}_{0})).
\end{equation}
Because of the identification of the integral orbits with the irreducible 
unitary representations this leads to a sum over the representations.
A special form of the character formula of Kirillov \cite{KI2} says that the symplectic volume 
of the co-adjoint orbit equals the dimension of the corresponding  
irreducible unitary representation. So the final form of the partition function is
\begin{equation}
Z=\sum_{\lambda} {\rm d}(\lambda)^{\chi(M)}\exp(-\frac{1}{\hbar}A_{M}C(\lambda)),
\end{equation}
where $\lambda$ denotes the irreducible unitary representation corresponding 
to the co-adjoint orbit, and ${\rm d}(\lambda)$ is the dimension of this 
representation. This is exactly the partition function for the two-dimensional 
Yang-Mills theory \cite{BT}. 
When we omit the Casimir term in the action we get just a sum over the dimensions of the 
representations, which is the correct result for the BF-theory, see e.g. \cite{BT2}.

\section{Conclusions and Outlook}

The Poisson-Sigma model is more than a unified framework for different topological and 
semi-topological field theories. Due to its reformulation of the degrees of freedom of the 
theories in terms of the coordinates of a Poisson manifold it achieves a description in terms 
of the natural variables of general dynamical systems. Gauge theories, which are characterized 
by singular Lagrangians, cannot in general be described in terms of symplectic manifolds; 
the foliation which is characteristic for Poisson manifolds is neccesary.

The advantages of such a description of these field theories is at least twofold. First, it 
allows one to discuss the quantization of the classical field theory by a direct application 
of the techniques of deformation quantization. Second, for integrable systems the general 
dynamical concepts of integrability may be utilized in order to reduce the partition function 
of the theory.

To some extent the above remarks are illustrated in the present work. The use of 
Casimir-Darboux coordinates allowed essential simplifications. In Section (4) we achieved a 
full reduction of the path integral in a special case through the use of concepts and 
theorems involving the symmetries of general dynamical sytems. 

We believe that further research will uncover ways of utilizing these structures 
even more thoroughly. The techniques used here should in principle be applicable in more 
general situations than the particular case we considered in Section (4). We also hope to be 
able to treat more general manifolds. This would allow in particular a more direct 
comparison with canonical quantization procedures. Finally, as already discussed in the 
Introduction, an understanding of the mechanisms active in the general case should allow the 
resolution of problems encountered in particular field theories.
\bigskip\\
{\large\bf Acknowledgement}

This work was supported in part (T. Schwarzweller) by the {\em Deutsche Forschungsgesellschaft} in 
connection with the Graduate College for Elementary Particle Physics in Dortmund.  
The work done by Beatrice Bucker and Tim Eynck in their Diplom theses was very helpful for 
the authors. We also thank Peter Henselder for a critical reading of the manuscript.


\begin{thebibliography}{9999}
\bibitem{SS1} P.Schaller, T.Strobl, Poisson Structure Induced (Topological)
Field Theories in Two Dimensions,
Mod. Phys. Lett. {\bf A9} (1994), 3129
\bibitem{SS2} P.Schaller, T.Strobl, Poisson-$\sigma$-models: A Generalization of 2d 
Gravity-Yang-Mills Systems, 
Talk delivered at the Conference on Integrable Systems, Dubna 1994, 
e-Print Archive: {\bf hep-th/9411163}
\bibitem{H} M.Henneaux, C.Teitelboim, Quantization of Gauge Systems, 
Princeton University Press, Princeton, New Jersey (1992)
\bibitem{BV} I.A.Batalin, G.A.Vilkovisky, Gauge Algebra and Quantization,
Phys. Lett. {\bf 69B} (1977), 309
\bibitem{CF} A.S.Cattaneo, G.Felder, A Path Integral Approach to the Kontsevich Quantization 
Formula, e-Print Archive: {\bf math/9902090}
\bibitem{KO}  M.Kontsevich, Deformation Quantization of Poisson Manifolds I,
e-Print Archive:\linebreak {\bf q-alg/9709040}
\bibitem{KU1} W.Kummer, H.Liebl, D.V.Vassilevich, Exact Path Integral Quantization of 
generic 2d Dilaton Gravity, Nucl. Phys. {\bf B493} (1997), 491
\bibitem{KU2}  W.Kummer, H.Liebl, D.V.Vassilevich, Integrating Geometry in general 2d 
Dilaton Gravity with Matter, Nucl. Phys. {\bf B544} (1999), 403
\bibitem{LP1} J.M.F.Labastida, E.P\`{e}rez, Kontsevich Integral for Vassiliev Invariants from 
Chern-Simons Pertubation Theory in the Light Cone Gauge, J. Math. Phys {\bf 39(10)} (1998), 5183
\bibitem{BOT} R. Bott, C. Taubes, On the Self Linking of Knots,
J. Math. Phys. {\bf 35(10)} (1994), 5247
\bibitem{HS} A.C.Hirshfeld, U.Sassenberg, Explicit Formulation of a third order finite Knot 
Invariant derived from Chern-Simons Theory,
Journal of Knot Theory and its Ramifications {\bf 5} (1996), 805
\bibitem{LP2} J.M.F.Labastida, E.P\`{e}rez, Combinatorial Formulation for Vassiliev 
Invariants from Chern-Simons Gauge Theory,
e-Print Archive: {\bf hep-th/9807155} (1998)
\bibitem{BT} A.Migdal, Recursion Relations in Gauge Theories, Zh. Eksper. Teoret. Fiz.
{\bf 69} (1975), 810 (Soviet Physics JETP. {\bf 42}, 413).\\
E.Witten, On Quantum Gauge Theories in Two Dimensions, Comm. Math. Phys. 
{\bf 141} (1991), 153\\
M.Blau, G.Thompson, Quantum Yang-Mills Theory on Arbitrary Surfaces,
Int. J. Mod. Phys. {\bf A7} (1992), 3781
\bibitem{Wi1} E.Witten, A Note on the Antibracket Formalism, 
Mod. Phys. Lett. {\bf A5} (1990), 487
\bibitem{AKSZ} M.Alexandrov, M.Kontsevich, A.Schwarz, O.Zaboronsky, The Geometry of 
the Master Equation and Topological Quantum Field Theory, 
Int. J. Mod. Phys. {\bf A12} (1997), 1405
\bibitem{GPS} J.Gomis, J.Paris, S.Stuart, Antibrackets, Antifields and Gauge-Theory 
Quantization, Phys. Rept. {\bf 259} (1995), 1
\bibitem{VA} I.Vaisman, Lectures on the Geometry of Poisson Manifolds, 
Progress in Mathematics Volume {\bf 118}, Birkh\"auser, Basel, 1994
\bibitem{WE} A.Weinstein, The local Structure of Poisson Manifolds, 
J. Differential Geometry {\bf 18} (1983), 523
\bibitem{KI1} A.Kirillov, The Orbit Method I, Geometric Quantization, 
in Representation Theory of Groups and Algebras, 
 Contemporary Mathematics Volume {\bf 145}, 1993
\bibitem{BT2}M.Blau, G.Thompson, Lectures on 2d Gauge Theories, 
presented at the 1993 Trieste Summer School in High Energy
Physics and Cosmology, \linebreak
e-Print Archive: {\bf hep-th/9310144}
\bibitem{KI2} A.Kirillov, Elements of the Theory of Representations, 
Grundlehren der mathematischen Wissenschaften {\bf 220},
Springer, Berlin, 1976
\end{thebibliography}
\end{document}